\UseRawInputEncoding
\documentclass[aps,prb,reprint,preprintnumbers,amsmath,amssymb,superscriptaddress, showpacs,showkeys,floatfix]{revtex4-2}

\usepackage{graphicx}
\usepackage{bm}
\usepackage{color} 
\usepackage[separate-uncertainty,multi-part-units = single ]{siunitx}
\usepackage{float}
\usepackage{graphicx}
\usepackage{dcolumn}
\usepackage{bm}
\usepackage{amsmath}
\usepackage{amssymb}
\usepackage{bbold}
\usepackage{siunitx}
\usepackage{color}
\usepackage{braket}
\usepackage{lipsum}
\usepackage{hyperref}
\usepackage{natbib}
\usepackage{miller}
\usepackage{xcolor}
\usepackage{tikz}
\usepackage{url}
\usepackage{physics}
            
\begin{document}

\title[]{Fabrication of low-loss Josephson parametric devices}

\author{K. E.~Honasoge}
\email[]{kedar.honasoge@wmi.badw.de}
\affiliation{Walther-Mei{\ss}ner-Institut, Bayerische Akademie der Wissenschaften, 85748 Garching, Germany}
\affiliation{School of Natural Sciences, Technische Universit\"{a}t M\"{u}nchen, 85748 Garching, Germany}

\author{M. Handschuh}
\affiliation{Walther-Mei{\ss}ner-Institut, Bayerische Akademie der Wissenschaften, 85748 Garching, Germany}
\affiliation{School of Natural Sciences, Technische Universit\"{a}t M\"{u}nchen, 85748 Garching, Germany}

\author{W. K. Yam}
\affiliation{Walther-Mei{\ss}ner-Institut, Bayerische Akademie der Wissenschaften, 85748 Garching, Germany}
\affiliation{School of Natural Sciences, Technische Universit\"{a}t M\"{u}nchen, 85748 Garching, Germany}

\author{S. Gandorfer}
\affiliation{Walther-Mei{\ss}ner-Institut, Bayerische Akademie der Wissenschaften, 85748 Garching, Germany}
\affiliation{School of Natural Sciences, Technische Universit\"{a}t M\"{u}nchen, 85748 Garching, Germany}

\author{D. Bazulin}
\affiliation{Walther-Mei{\ss}ner-Institut, Bayerische Akademie der Wissenschaften, 85748 Garching, Germany}
\affiliation{School of Natural Sciences, Technische Universit\"{a}t M\"{u}nchen, 85748 Garching, Germany}

\author{N. Bruckmoser}
\affiliation{Walther-Mei{\ss}ner-Institut, Bayerische Akademie der Wissenschaften, 85748 Garching, Germany}
\affiliation{School of Natural Sciences, Technische Universit\"{a}t M\"{u}nchen, 85748 Garching, Germany}

\author{L. Koch}
\affiliation{Walther-Mei{\ss}ner-Institut, Bayerische Akademie der Wissenschaften, 85748 Garching, Germany}
\affiliation{School of Natural Sciences, Technische Universit\"{a}t M\"{u}nchen, 85748 Garching, Germany}

\author{A. Marx}
\affiliation{Walther-Mei{\ss}ner-Institut, Bayerische Akademie der Wissenschaften, 85748 Garching, Germany}

\author{R. Gross}
\affiliation{Walther-Mei{\ss}ner-Institut, Bayerische Akademie der Wissenschaften, 85748 Garching, Germany}
\affiliation{School of Natural Sciences, Technische Universit\"{a}t M\"{u}nchen, 85748 Garching, Germany}
\affiliation{Munich Center for Quantum Science and Technology (MCQST), 80799 Munich, Germany}

\author{K. G. Fedorov}
\email{kirill.fedorov@wmi.badw.de}
\affiliation{Walther-Mei{\ss}ner-Institut, Bayerische Akademie der Wissenschaften, 85748 Garching, Germany}
\affiliation{School of Natural Sciences, Technische Universit\"{a}t M\"{u}nchen, 85748 Garching, Germany}
\affiliation{Munich Center for Quantum Science and Technology (MCQST), 80799 Munich, Germany}


\begin{abstract}
Superconducting circuits incorporating Josephson elements represent a promising hardware platform for quantum technologies. Potential applications include scalable quantum computing, microwave quantum networks, and quantum-limited amplifiers. However,  progress in Josephson junction-based quantum technologies is facing the ongoing challenge of minimizing loss channels. This is also true for parametric superconducting devices based on nonlinear Josephson resonators. In this work, we report on the fabrication and characterization of low-loss Josephson parametric devices operated in the GHz frequency range, showing record internal quality factors. Specifically, we achieve internal quality factors $Q_\mathrm{int}$ significantly exceeding $10^5$ for both Josephson parametric converters and Josephson parametric amplifiers in the single-photon regime by fitting the scattering data. We confirm the extracted $Q_\mathrm{int}$ values by analyzing purity of squeezed vacuum states generated by these devices. These low-loss devices mark a significant step forward in realizing high-performance quantum circuits, enabling further advancements in superconducting quantum technologies.
\end{abstract}

\maketitle

\section{Introduction}

Superconducting circuits based on Josephson junctions have become a cornerstone in modern, solid-state quantum technologies. These circuits are key for realizing scalable quantum computing, enabling secure quantum communication, and probing fundamental new physics, e.g. in dark matter detection \cite{aruteQuantumSupremacyUsing2019,abdoJosephsonAmplifierQubit2011,bartramDarkMatterAxion2023,backesQuantumEnhancedSearch2021} or quantum phase transitions \cite{Chen2023, DiCandia2023}. Superconducting circuits also serve as the basis for a variety of novel experiments with propagating quantum microwave signals. These experiments, which exploit two-mode squeezed vacuum states, include quantum teleportation, quantum key distribution, and quantum illumination \cite{fedorovExperimentalQuantumTeleportation2021,fesquetDemonstrationMicrowaveSingleshot2024,assoulyQuantumAdvantageMicrowave2023, Kronowetter1, Kronowetter2}. Josephson parametric amplifiers (JPAs) and converters (JPCs) are central components in these experiments, providing the toolbox for the generation and manipulation of propagating squeezed microwave states \cite{Cryolink, fedorovFinitetimeQuantumEntanglement2018b}. Last but not least, various versions of JPAs are indispensable for the fast single-shot readout of superconducting qubits \cite{touzardGatedConditionalDisplacement2019,sunadaFastReadoutReset2022a}. 

The performance of Josephson parametric devices is often limited by the presence of various mechanisms leading to energy dissipation and decoherence. In superconducting quantum circuits several such loss mechanisms have been identified, including two-level systems (TLSs), surface spin states, non-equilibrium quasiparticles, radiative losses, Abrikosov vortices, and even irradiation by cosmic high-energy particles \cite{goetzLossMechanismsSuperconducting2016,mcraeMaterialsLossMeasurements2020,mullerUnderstandingTwolevelsystemsAmorphous2019c,chayanunCharacterizationProcessrelatedInterfacial2024b,wangCandidateSourceFlux2015,braumullerCharacterizingOptimizingQubit2020d,kumarOriginReduction12016,vandammeArgonMillingInducedDecoherenceMechanisms2023a}. 
These losses not only deteriorate the performance of parametric devices by adding extra noise photons to the signal but also reduce the coherence time of superconducting qubits, which are based on similar material systems and fabrication techniques. For resonators, losses are usually quantified by the internal quality factor, $Q_{\mathrm{int}}$, defined as the ratio of stored energy to the average amount of energy lost per cycle via internal loss channels. Reducing internal losses or, equivalently, increasing the internal quality factor of superconducting devices is crucial for advancing superconducting quantum technologies.

\begin{figure*}[!t]
	\begin{center}
		\includegraphics[width=\linewidth,angle=0,clip]{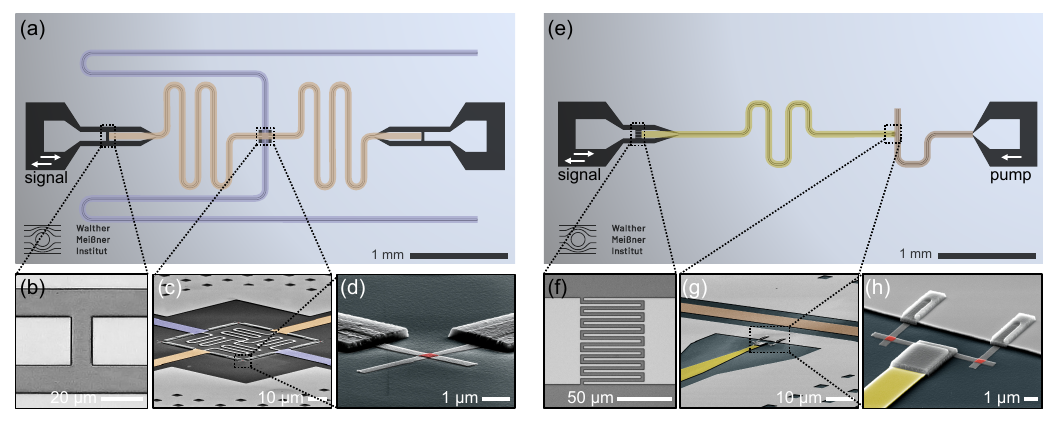}
	\end{center}
    \caption{Design renderings and experimental images of  fabricated key components of the Josephson parametric converter (JPC) and Josephson parametric amplifier (JPA) with false-color overlays. (a) The JPC features two $\lambda / 2$ coplanar waveguide (CPW) resonators (blue and yellowish-brown). The yellowish-brown resonator is coupled to the input and output ports via a capacitor, shown in panel (b). The JPC resonators are coupled to each other via an inductively shunted Josephson ring modulator (JRM), shown in panel (c). The JRM consists of four Manhattan-style $\mathrm{Al/AlO_x/Al}$ Josephson junctions (JJs), shown in panel (d), at the vertices of a rectangular loop with Nb shunts. (e) The JPA features a $\lambda/4$ CPW resonator (yellow), coupled to an input port by an interdigitated capacitor, shown in panel (f). The resonator is short-circuited to ground via a dc-SQUID, formed by two Manhattan-style $\mathrm{Al/AlO_x/Al}$ JJs in a rectangular loop, illustrated in panels (g) and (h). An additional transmission line (brown), referred to as the pump line in the main text, is inductively coupled to the dc-SQUID. The JJs in panels (d) and (h) are highlighted in red.}
	\label{Fig:Fig1}
\end{figure*}
Over the past two decades, efforts to reduce losses in superconducting circuits have led to substantial progress. Fixed-frequency superconducting qubits with energy relaxation times $T_{\mathrm{1}}$ approaching \SI{1}{\milli\second} have been achieved \cite{tuokkolaMethodsAchieveNearmillisecond2024a,somoroffMillisecondCoherenceSuperconducting2023}. These advances have been driven by improvements in material quality and selection, fabrication processes, surface treatments, and circuit design. They allow for a reduction of both the amount of loss channels and their coupling to the critical components of superconducting quantum circuits.  However, consistently achieving low internal loss rates in tunable Josephson junction-based circuits operated at a low signal power corresponding to the single photon level remains a significant challenge \cite{pogorzalekHystereticFluxResponse2017a,rochWidelyTunableNondegenerate2012,hutchingsTunableSuperconductingQubits2017a,abdurakhimovLonglivedCapacitivelyShunted2019,yanFluxQubitRevisited2016,chavez-garciaWeaklyFluxTunableSuperconducting2022}. The coherence times and internal quality factors of frequency-tunable qubits and resonators are considerably worse than their fixed-frequency counterparts. This has led to extensive research to understand and identify microscopic origins of additional losses in flux-tunable devices  \cite{wangCandidateSourceFlux2015,braumullerCharacterizingOptimizingQubit2020d,kumarOriginReduction12016}.

Here, we report on the fabrication and characterization of two types of Josephson parametric devices: a JPC and a JPA. These devices are based on a Josephson ring modulator (JRM) and a direct current Superconducting Quantum Interference Device (dc-SQUID), respectively. By combining established techniques and optimizing our fabrication process by specific measures, such as surface treatment prior to Josephson junction fabrication and refining argon ($\mathrm{Ar}^{+}$) ion milling, we succeed in realizing JPAs and JPCs with internal quality factors, $Q_{\mathrm{int}}$, exceeding $10^5$ at the single-photon power level. This represents a significant improvement over previously reported values and marks an important step toward low-noise Josephson parametric devices and tunable circuits. Our findings indicate that significant improvements of intrinsic losses in conventional Josephson superconducting circuits are possible with an improved surface treatment.


\section{Circuit design}

The designs of our Josephson parametric devices, the JPC and JPA, are illustrated in Figs.\,\ref{Fig:Fig1}(a) and (e), respectively. These parametric devices consist of coplanar waveguide (CPW) resonators coupled to a nonlinear, flux-tunable Josephson circuit. The CPW resonators are designed using electromagnetic finite element simulations in \textsc{Ansys HFSS}. The width of the central electrode and its separation to the ground are set to $\SI{10}{\micro\meter}$ and $\SI{6}{\micro\meter}$, respectively, resulting in a characteristic impedance of $Z_0 \approx \SI{50}{\ohm}$. The devices are fabricated using niobium (Nb) and aluminum (Al) through a multi-step process. The CPW resonators consist of Nb thin films that have been deposited by magnetron sputtering on high-resistivity silicon (Si) substrates and patterned by reactive ion etching. The Josephson circuits in both devices are based on $\mathrm{Al/AlO_x/Al}$ Josephson junctions (JJs). The JPC is based on a JRM containing four JJs, and the JPA includes a dc-SQUID with two JJs. Details of the complete fabrication process are provided later in Sec.\,\ref{expov}. The equivalent circuits for the JPC and JPA are shown in Figs.\,\ref{Fig:Fig3}(c) and (d).

The JPC features two half-wavelength ($\lambda^{A,B}/2$) CPW resonators, labeled \textit{A} and \textit{B} in Fig.\,\ref{Fig:Fig3}(c) [colored yellowish-brown and blue, respectively, in Fig.\,\ref{Fig:Fig1}(a)] with bare resonator frequencies of $\omega_r^A/2\pi \approx \SI{5.2}{\GHz}$ and $\omega_r^B/2\pi \approx \SI{7.2}{\GHz}$. These resonators are coupled via the inductively shunted JRM positioned at the center of both resonators, as shown in Fig.\,\ref{Fig:Fig1}(a). The JRM comprises a rectangular Nb loop with intersecting inductive shunts, where four identical Josephson junctions are positioned at the loop vertices [cf.~Fig.\,\ref{Fig:Fig1}(c)]. Two symmetric gap capacitors with capacitance $C_\mathrm{c}^A \approx \SI{7}{\fF}$  couple resonator \textit{A} to the input and output ports [cf.~Fig.\,\ref{Fig:Fig1}(b)]. Resonator \textit{B} is decoupled from the external microwave circuit and only weakly accessible through resonator \textit{A}. As such, resonator \textit{B} can serve as a quantum memory in future experiments.

A finite coupling of resonator \textit{A} to the JRM allows for tuning its resonance frequency by applying an external magnetic flux, $\Phi_{\mathrm{ext}}$. This frequency tunability is described by \cite{rochWidelyTunableNondegenerate2012},
\begin{equation}
    \label{jpc_fd}
    \omega_{\mathrm{JPC}}^{A} = \omega_r^{A}  \frac{\pi^2L_{A}^{\lambda/2}/2}{\pi^2L_{A}^{\lambda/2}/2 + L_{A}( \varphi_{\mathrm{ext}} )} ,
\end{equation}
where $\omega_r^A$ is the bare resonator frequency, $L_A^{\lambda/2} = 2Z_0/(\pi\omega_r^A)$ is its lumped-element equivalent inductance, and $\varphi_{\mathrm{ext}} = \tfrac{1}{4}(2\pi\Phi_{\mathrm{ext}}/ \Phi_0)$ is the normalized flux threading each of the four loops of the JRM, with the flux quantum $\Phi_0 = h/2e$. Using the reduced flux quantum $\phi_0 = \Phi_0 / 2\pi$, the nonlinear flux-tunable inductance of the JRM can be expressed as
\begin{equation}
    \label{jrm_inductance}
        L_A( \varphi_{\mathrm{ext}}) =  \phi_0 ^2 \left( \frac{E_L}{2} + E_J \cos \left(\varphi_{\mathrm{ext}}\right)  \right)^{-1}
\end{equation}
for $E_L/4 + E_J \cos(\varphi_{\mathrm{ext}}) > 0$. Here, $E_L = \phi_0^2 / L$ is the energy associated with the flux stored in the internal shunt inductances $L$, and $E_J = \phi_0^2/L^0_J$ the Josephson energy of each junction having a zero-flux Josephson inductance of $L^0_J$.

The JPA features a $\lambda^R/4$ CPW resonator terminated to ground via a dc-SQUID. An image of the JPA and its equivalent circuit are shown in Figs.\,\ref{Fig:Fig1}(e) and (d), respectively. The JPA resonator has a bare resonance frequency of $\omega_r/2\pi \approx \SI{6.1}{\GHz}$ and is coupled to the input signal port through an interdigitated capacitor, as shown in Fig.\,\ref{Fig:Fig1}(f), with the coupling capacitance of $C_\mathrm{c} \approx \SI{50}{\femto \F}$. Another CPW line accessible via the pump port on the opposite part of the chip [cf.~Fig.\,\ref{Fig:Fig1}(e)] is inductively coupled to the dc-SQUID and serves for flux-pumping of the JPA. Two $\mathrm{Al/AlO_x/Al}$ JJs embedded into a rectangular superconducting loop form the dc-SQUID [see Fig.\,\ref{Fig:Fig1}(g)]. The dc-SQUID provides a flux-tunable nonlinear inductance, allowing the JPA resonance frequency, $\omega_{\mathrm{JPA}}$, to be tuned with an external magnetic flux, $\Phi_{\mathrm{ext}}$. The flux-dependent JPA frequency can be expressed using a distributed-element model as \cite{wustmannParametricResonanceTunable2013b, pogorzalekHystereticFluxResponse2017a, wallquistSelectiveCouplingSuperconducting2006a}
\begin{equation}
\label{jpa_fd}
\omega_{\mathrm{JPA}}(\Phi_{\mathrm{ext}}) = \omega_r \left( 1 + \frac{L_S(\Phi_{\mathrm{ext}}) + L_{\mathrm{loop}/4}}{L_r}\right)^{-1},
\end{equation}
where $\omega_r$ is the bare resonance frequency, $L_r$ the inductance of the resonator, $L_{\mathrm{loop}}$ the geometric inductance of the SQUID loop, and $L_S$ the tunable dc-SQUID inductance. This tunable inductance is given by \cite{sandbergTuningFieldMicrowave2008a}
\begin{equation}
\label{squidInd}
    L_S(\Phi_{\mathrm{ext}}) = \Phi_0 / 4\pi I_c|\cos(\pi \Phi_{\mathrm{ext}}/\Phi_0)|,
\end{equation}
where $I_c$ is the critical current of a single JJ in the dc-SQUID. 

\section{Experimental Techniques}
\label{expov}
In this section, we provide an overview of the methods and techniques used to fabricate and characterize our Josephson parametric devices. We first provide a step-by-step description of our fabrication process allowing us to achieve the low-loss characteristics of our devices. We then discuss the cryogenic measurement setup used for the measurement of the internal quality factor.

\subsection{Sample Fabrication}

\begin{figure}[!t]
	\begin{center}
		\includegraphics[width=\linewidth,angle=0,clip]{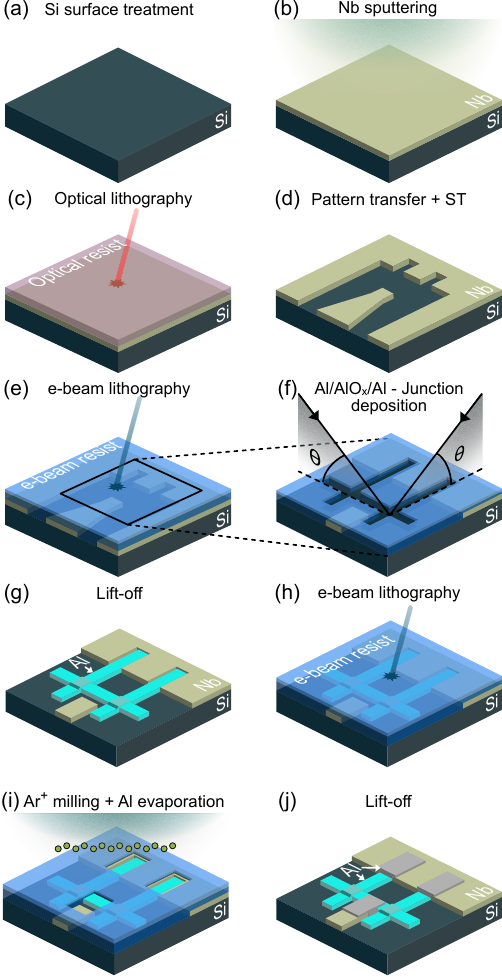}
	\end{center}
	\caption{Fabrication procedure for Josephson parametric devices. (a,b) The process starts with the surface treatment of the high-resistivity Si substrates, followed by sputter deposition of the Nb film in an ultra-high vacuum system to ensure clean metal-substrate interfaces. (c,d) The Nb base layer is patterned using optical lithography and reactive ion etching. A subsequent surface treatment ensures clean metal-air and substrate-air interfaces. (e-g) Josephson junctions are fabricated via electron beam (e-beam) lithography using a CSAR/PMMA double-resist stack, followed by double-angle Al evaporation with intermediate oxidation and lift-off steps. (h-j) Metallized Al and Nb layers are connected by Al bandages, realized by double-resist e-beam lithography, Al evaporation, and lift-off, with an intermediate Ar$^+$ ion milling step to ensure clean metal-metal interfaces. }
	\label{Fig:Fig2}
\end{figure}

The main steps of our fabrication process are illustrated in Fig.\,\ref{Fig:Fig2}. It starts with acetone and isopropyl alcohol (IPA) pre-cleaning of the \SI{545}{\micro\meter} thick, \hkl<100> oriented high resistivity ($\geq$\SI{10}{\kilo\ohm\cm}) silicon (Si) substrates. Next, in order to ensure a clean metal-substrate interface, the substrates are treated in a piranha solution ($\mathrm{H_2SO_4:H_2O_2}=3:1$) for \SI{10}{\minute} at \SI{80}{\degreeCelsius} followed by rinsing in de-ionized water for \SI{2}{\minute}. This step ensures removal of organic contaminants and particulates \cite{placeNewMaterialPlatform2021a}. Native oxides are removed by subsequently treating the substrate with a buffered oxide etch (BOE) solution for \SI{30}{\second} \cite{altoeLocalizationMitigationLoss2022b}. The BOE-water solution consists of $\mathrm{6.5\,wt\%}$ hydrofluoric acid ($\mathrm{HF}$) and $\mathrm{34.8 \,wt\%}$ ammonium fluoride ($\mathrm{NH_4F}$). Immediately after cleaning, the substrates are inserted into a load-locked ultra-high vacuum deposition system (\textsc{Plassys  MEB550 S4-I}) with separate niobium (Nb) sputtering and aluminum (Al) evaporation chambers. First, the surface is metallized with $\thicksim$\SI{150}{\nm} of Nb by dc-magnetron sputtering at room temperature. Large design elements (with dimensions $>$ \SI{1}{\micro\meter}) are patterned into the Nb film using optical lithography, using the \textsc{AZ-701} optical resist and \textsc{PicoMaster 200} laser writer. The substrates are then developed in \textsc{AZ-726}-MIF (metal ion free), rinsed with de-ionized water and dry etched in an inductively coupled reactive-ion etcher (\textsc{Oxford Instruments PlasmaLab 80 Plus}) with sulfur hexafluoride ($\mathrm{SF_6}$) for pattern transfer. After the etching step, the residual resist is stripped off with \textsc{TechniStrip-P1331} at \SI{90}{\degreeCelsius} for \SI{5}{\minute}, followed by a rinsing step with de-ionized water and IPA. 

After patterning the base Nb layer, the substrates are treated in the BOE solution for \SI{20}{\minute}. This step removes any process oxides of Si and Nb formed during the fabrication process so far \cite{altoeLocalizationMitigationLoss2022b} and ensures a clean metal-air and substrate-air interfaces. Next, the Josephson junctions are defined in a double-layer resist stack consisting of PMMA (polymethyl methacrylate) as the top layer and its co-polymer CSAR (poly-$\alpha$-methylstyrene-co-$\alpha$-chloroacrylate methylester) as bottom layer, using electron beam (e-beam) lithography. After development in resist-specific developers, \textsc{AR600-56} and \textsc{AR600-546}, the substrates are treated with BOE for \SI{30}{\second} to ensure a clean metal-substrate interface. The substrates are immediately loaded into the loadlock for double-angle aluminum (Al) evaporation with an intermediate oxidation step, forming the Manhattan-style JJs. Following the lift-off of the residual Al, the substrate is spin-coated for e-beam lithography with the double-layer resist to define a ''bandage" layer. The bandage involves a final Al evaporation after argon ion ($\mathrm{Ar}^+$) milling. The $\mathrm{Ar}^+$ milling step ensures a clean metal-metal interface and a superconducting contact between the Nb and Al layers. Following the final lift-off and wire bonding, the devices are mounted into the dilution refrigerator for cryogenic characterization.

\subsection{Sample Characterization}
\begin{figure}[tb]
	\begin{center}
		\includegraphics[width=\linewidth,angle=0,clip]{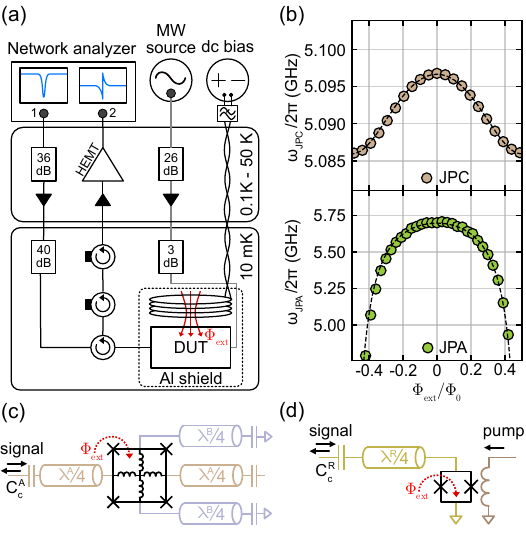}
	\end{center}
	\caption{Experimental setup, measured flux dependencies of resonance frequencies, and equivalent circuits of the realized Josephson parametric devices. (a) The JPCs and JPAs are enclosed in an Al shielding box (dotted line) and measured in reflection in a dilution refrigerator using a vector network analyzer. (b) Measured resonance frequencies of the JPC ($\omega_{\mathrm{JPC}}$) and JPA ($\omega_{\mathrm{JPA}}$) as a function of the flux bias. Colored circles represent data and dashed lines are fits. (c),(d) Equivalent circuits of the JPC and JPA, respectively.}
	\label{Fig:Fig3}
\end{figure}

Our Josephson parametric devices are characterized using cryogenic microwave measurements in a dilution refrigerator at temperatures close to its base temperature of about $T\thicksim\SI{10}{\milli\kelvin}$. A schematic of our experimental setup is shown in Fig.\,\ref{Fig:Fig3}(a). The device under test (DUT), either the JPC or JPA, is wire-bonded to a gold-plated printed circuit board (PCB) with CPW transmission lines connected to SubMiniature Version A (SMA) connectors. The DUT and PCB are mounted in a gold-plated copper sample holder, which is placed inside of a superconducting Al magnetic shield and thermally anchored to the mixing chamber plate. An external superconducting magnetic coil mounted on the sample holder provides the magnetic flux bias to the Josephson parametric devices. The sample holder and magnetic coil are further thermalized to the mixing chamber plate with silver ribbons.

Single-port reflection measurements on the DUT are obtained by connecting the sample box to a microwave cryogenic circulator through a superconducting coaxial cable [cf.~Fig.\,\ref{Fig:Fig3}(a)]. For the JPA measurements, a second input copper coaxial line is connected to the pump port to apply a pump tone provided by a microwave source. The circulator input lines for the JPC and JPA have a total cryogenic attenuation of \SI{66}{\dB} and \SI{76}{\dB}, respectively, while the JPA pump port input line has \SI{29}{\dB} of cryogenic attenuation. The circulator output lines pass through a pair of cryogenic microwave isolators and are connected via superconducting coaxial cables to a high-electron-mobility transistor (HEMT) amplifier and an additional room-temperature amplifier. The magnetic field coil is connected to a room-temperature current source via normal- and superconducting wires, filtered with low-pass filters. Spectroscopic measurements are performed by measuring the complex scattering coefficient $S_{\mathrm{21}}$ as a function of frequency, power, and applied magnetic flux using a four-port Rohde \& Schwarz vector network analyzer (VNA). We label the measured scattering coefficient as $S_{\mathrm{21}}$ for convenience, as the VNA serves as the reference point; in other contexts, the same coefficient referenced to the DUT is often labeled $S_{\mathrm{11}}$.

\section{Results}

We first discuss the dependence of the resonance frequencies of our Josephson parametric devices on the applied magnetic flux. Measuring this flux dependence allows us to estimate the critical current $I_{\mathrm{c}}$ of respective JJs. Since both the JPC and JPA contain Josephson elements acting as flux-tunable inductors, we can tune their resonance frequencies by adjusting the flux bias $\Phi_{\mathrm{ext}}$ through the JRM and dc-SQUID, respectively. Experimentally, we determine the resonance frequency by measuring $S_{\mathrm{21}}$ as a function of the probe frequency and dc current through the superconducting magnetic field coil. The flux-dependent resonance frequencies, extracted from the real part of the scattering coefficient, are shown in Fig.\,\ref{Fig:Fig3}(b). 

\begin{figure*}[!ht]
    \centering
		\includegraphics[width=\linewidth,angle=0,clip]{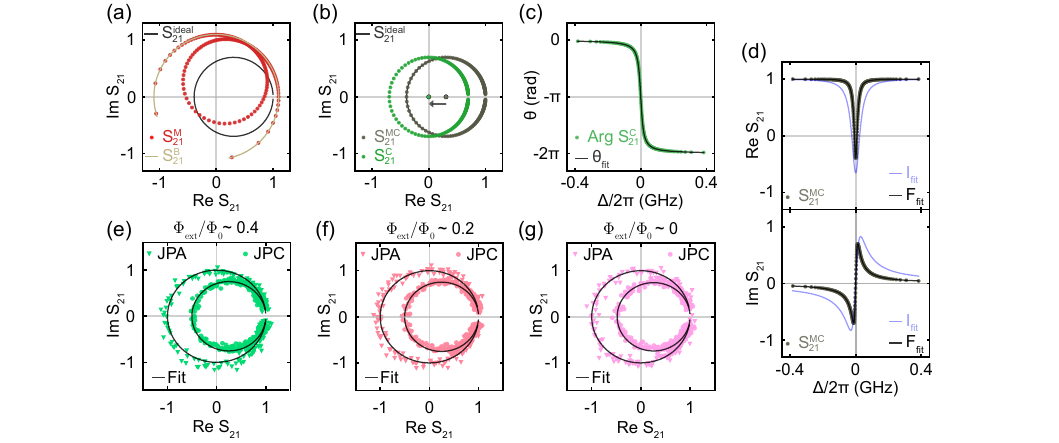}
	\caption{Analysis of the internal quality factors of Josephson parametric devices. (a) By an additional background contribution $S_{21}^{\mathrm{B}}$ the ideal intrinsic scattering data $S_{21}^{\mathrm{ideal}}$ is transformed yielding $S_{21}^{\mathrm{M}}$ corresponding to the actually measured scattering data. (b) The background corrected scattering parameter $S_{21}^{\mathrm{MC}}$ can be obtained by methods described in the main text. $S_{21}^{\mathrm{C}}$ is obtained by centering $S_{21}^{\mathrm{MC}}$ on the complex plane. (c) A phase vs frequency fit to $S_{21}^{\mathrm{C}}$ yields the loaded quality factor ${Q_{\mathrm{l}}}$. (d) A least-squares minimization fit of Eq.~(\ref{ioref}) to the real and imaginary part of $S_{21}^{\mathrm{MC}}$ leads to an estimate of the external quality factor $Q_{\mathrm{ext}}$. The initial fits are shown in blue ($\mathrm{I}_\mathrm{fit}$) and the final minimized fits in black ($\mathrm{F}_\mathrm{fit}$). (e-g) The same fitting procedure applied to the scattering data $S_{21}^{\mathrm{MC}}$ measured at the three flux points $\Phi_{\mathrm{ext}}/\Phi_{0} = 0, 0.2$, and $0.4$.}
	\label{Fig:Fig4}
\end{figure*}

Fitting the measured $\omega_{\mathrm{JPC}}^A$ vs. $\Phi_\text{ext}$ data shown in Fig.\,\ref{Fig:Fig3}(b), assuming an equal flux distribution between the four JRM subloops, with Eq.~(\ref{jpc_fd}) yields the zero-flux resonance frequency $\omega_{\mathrm{JPC}}^A / 2\pi \approx \SI{5.17}{\giga\hertz}$ and a tunability range of about $\SI{10.8}{\mega\hertz}$. The extracted critical currents of the JJs in the JRM is obtained to be $I_{\mathrm{c}}^{\mathrm{JPC}} \approx 0.286 \pm \SI{0.023}{\micro\ampere}$, corresponding to the Josephson energy  $E_{\mathrm{J}}^{\mathrm{JPC}} /h\approx \SI{144}{\giga\hertz}$ for each junction. Additionally, we estimate the geometric internal shunt inductance of the JRM to be $L \approx \SI{61.8}{\nano\henry}$. Similarly, fitting the $\omega_{\mathrm{JPA}}(\Phi_\text{ext})$ data with Eq.~(\ref{jpa_fd}) leads to the zero flux resonance frequency of $\omega_{\mathrm{JPA}}/2\pi \approx \SI{5.71}{\giga\hertz}$ and a tunability range of $>\SI{750}{\mega\hertz}$. The extracted critical current of the JJs in the dc-SQUIDs is $I_c \approx 1.380 \pm \SI{0.031}{\micro \ampere}$, corresponding to a Josephson energy of $E_{\mathrm{J}}^{\mathrm{JPA}} /h\approx \SI{685}{\giga\hertz}$ for each junction. The dc-SQUID loop is estimated to have the geometric inductance of around $\SI{7.9}{\pico \henry}$.

Next, we focus on analyzing the microwave losses in the investigated JPC and JPA devices. Their response close to the resonance frequency $\omega_0$ is determined by losses that can be attributed to two different loss channels: (i) external losses arising from the finite coupling to the external microwave circuit and (ii) internal losses due to the coupling to various intrinsic loss channels.  External losses are quantified by the external quality factor, $Q_{\mathrm{ext}} = \omega_0/ \kappa_{\mathrm{ext}}$, with the loss rate, $\kappa_{\mathrm{ext}}$, which can be varied by changing the coupling capacitance, $C_{c}$, of the devices. Internal losses are quantified by $Q_{\mathrm{int}} = \omega_0/ \kappa_{\mathrm{int}}$, with the internal loss rate, $\kappa_{\mathrm{int}}$. The total loaded quality factor, $Q_{\mathrm{l}}$, is given by
\begin{equation}
\label{Qs}
    Q_{\mathrm{l}}^{-1} = Q_{\mathrm{ext}}^{-1} + Q_{\mathrm{int}}^{-1}.
\end{equation}

In order to model the Josephson parametric devices we consider them as driven dissipative harmonic oscillators. The scattering coefficients for a resonator with a total loss rate $\kappa = \kappa_{\mathrm{ext}} + \kappa_{\mathrm{int}}$ can be derived by applying standard quantum input-output formalism (see Ref.\,\cite{steck2007quantum,petersonParametricCouplingMicrowaves2020} for a detailed derivation). For a resonator probed in reflection at a single port through a perfect microwave circulator, the ideal scattering coefficient $S_{21}^\mathrm{ideal}$ is given by

\begin{equation}
\label{ioref}
S_{21}^{\mathrm{ideal}} = 1 + \frac{\kappa_{\mathrm{ext}}}{i \Delta - \kappa/2} \;.
\end{equation}
Here, $\Delta = \omega - \omega_0$ is the detuning between the probe frequency, $\omega$, and the resonance frequency, $\omega_0$. Near resonance, $\Delta \simeq 0$, the scattering coefficient $S_{21}^\mathrm{ideal}$ describes a circle in the complex plane, shown as the black line in Fig.\,\ref{Fig:Fig4}(a).

In reality, the ideal circle is distorted by a frequency-dependent microwave background, $S_{21}^\mathrm{B}$, of the experimental setup. This distortion includes amplitude scaling by attenuation (amplification) in the input (output) lines and a frequency-dependent phase delay, causing the circle in the complex plane to distort into an open loop. Additionally, a finite impedance mismatch in the measurement setup results in a tilt of the circle, while a Fano interference displaces it. As illustrated in Fig.\,\ref{Fig:Fig4}(a), these artifacts contribute to the measured scattering coefficient $S_{21}^\mathrm{M}$, which in addition to $S_{21}^\mathrm{ideal}$ also contains the background contribution $S_{21}^\mathrm{B}$. Finally, further setup-related features may imprint their signatures on the measured scattering coefficients, further increasing the complexity for their accurate analysis.

In practice, the microwave background arising from the above mentioned artifacts, fortunately, can be calibrated out \cite{stanleyValidatingParameterMeasurements2022,ranzaniTwoportMicrowaveCalibration2013}. This, however, imposes tedious experimental requirements. To address these challenges, robust fitting and measurement protocols have been developed \cite{probstEfficientRobustAnalysis2015a, khalilAnalysisMethodAsymmetric2012b, baityCircleFitOptimization2024a, riegerFanoInterferenceMicrowave2023}. These protocols incorporate data pre-processing techniques to correct for some distortions. Furthermore, one measures the scattering coefficients with unequal frequency spacing near the resonance frequency. This non-equidistant data point distributions improves parameter extraction accuracy as the system response near resonance is captured with an increased density of data points. In this work, we combine these techniques along with background normalization to recover the undistorted scattering parameters. Then, we fit the corrected experimental data with Eq.~(\ref{ioref}) to obtain the characteristic circuit parameters including the internal quality factor.

To determine the quality factors of the JPC, we measure the scattering coefficient $S_{21}^\mathrm{M}(\Delta)$ around the resonance frequency $\omega_{\mathrm{JPC}}^A$ with non-equidistant frequency spacing [cf.~Fig.\,\ref{Fig:Fig3}(b)] for different values of $\Phi_{\mathrm{ext}}$ and the drive power. To correct for the distortions and recover the undistorted scattering coefficients, we follow the approach of Ref.\,\cite{probstEfficientRobustAnalysis2015a}. First, we estimate the cable delay $\tau$ by using a linear fit to the phase angle ($\angle S_{21}^\mathrm{M}$) of the measured scattering coefficient.  This is followed by the minimization of a circle to the measured $S_{21}^\mathrm{M}$ data in the complex plane, solely parameterized by $\tau$. This allows us to get an estimate for any residual cable delay and correcting for it by performing a complex division of $S_{21}^\mathrm{M}$ by $\text{e}^{-i\omega\tau}$. After correcting for the cable delay, we can correct for scaling distortions caused by attenuation and amplification in the measurement lines. For this procedure an off-resonant operation point is used. Additionally, we correct for rotations in the complex plane resulting from impedance mismatch by referencing the center of the measured resonance circle. According to Eq.~(\ref{ioref}), the off-resonant point ($\Delta \gg \kappa$) ideally lies at $(1,0)$ in the complex plane and the center of the resonance circle on the real axis. By normalizing the measured data to that obtained for the off-resonant operation point and the angle between the real axis and the center of the resonance circle, we finally get the corrected scattering data, $S_{21}^\mathrm{MC}$.

Finally, after these correction steps, the near-ideal circle $S_{21}^\mathrm{MC}$ is shifted to the center of the complex plane to obtain $S_{21}^\mathrm{C}$. The total loss rate $\kappa$ can be derived from $S_{21}^\mathrm{C}$ by fitting the phase angle $\theta$ vs. detuning $\Delta$ data to the expression \cite{probstEfficientRobustAnalysis2015a,gaoPhysicsSuperconductingMicrowave2008}

\begin{equation}
    \label{pvsf}
    \theta(\Delta) = \theta_0 + 2 \arctan \left(  -\frac{2\Delta}{\kappa}  \right) \: .
\end{equation}
Here, $\theta$ and $\theta_0$ are the phase angle ($\angle S_{21}^\mathrm{C}$)  and residual phase offset [see Fig.\,\ref{Fig:Fig4}(c)].

\begin{figure}[tb]
	\begin{center}
		\includegraphics[width=\linewidth,angle=0, clip]{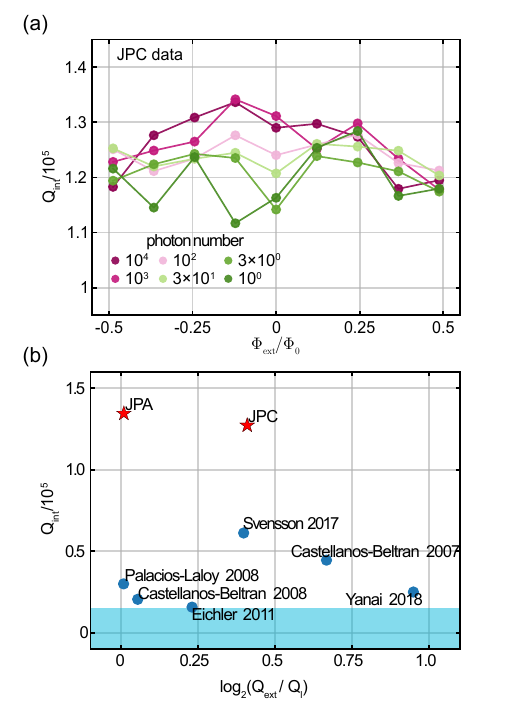}
	\end{center}
	\caption{(a) Internal JPC quality factors $Q_{\mathrm{int}}$ as a function of applied flux and input signal power given as the average number of photons in the resonator, $\langle n\rangle$. (b) Comparison of $Q_{\mathrm{int}}$ values of various Josephson parametric devices with different  coupling efficiencies, $Q_{\mathrm{ext}} / Q_{\mathrm{l}}$. The red stars refer to $Q_{\mathrm{int}}$ of the JPC and JPA devices from this work. Blue circles denote values reported in the literature \cite{abdoNondegenerateThreewaveMixing2013,castellanos-beltranAmplificationSqueezingQuantum2008d,castellanos-beltranWidelyTunableParametric2007, eichlerObservationTwoModeSqueezing2011,flurinJosephsonMixerSwiss2014,krantzInvestigationNonlinearEffects2013b,liFluxCoupledTunable2023a, palacios-laloyTunableResonatorsQuantum2008b,rochWidelyTunableNondegenerate2012,simoenCharacterizationMultimodeCoplanar2015,svenssonPeriodtriplingSubharmonicOscillations2017,uhlFluxtunableYBa2Cu3O7Quantum2023a,wangTunableSuperconductingResonators2024,xuMagneticFieldResilientQuantumLimited2023a,yanaiCoherenceNonlinearityMechanical2018c}. The blue-shaded area marks the region of $Q_{\mathrm{int}} < 1.5 \times 10^4$, where $Q_{\mathrm{int}}$ values are omitted for clarity.   }
	\label{Fig:Fig5}
\end{figure}

After having determined the total loss rate $\kappa$, we still need to determine  $\kappa_{\mathrm{ext}}$ to obtain the internal loss rate $\kappa_{\mathrm{int}}= \kappa - \kappa_{\mathrm{ext}}$. To derive $\kappa_{\mathrm{ext}}$, we perform  least square fits of the real and imaginary components of $S_{21}^\mathrm{MC}$. As shown in the upper and lower panel of Fig.\,\ref{Fig:Fig4}(d), the fitting procedure is initialized ($\mathrm{I}_{\mathrm{fit}}$) with the previously determined parameters. Using only $\kappa_{\mathrm{ext}}$ as a fitting parameter, the least-squares minimization yields the final fit ($\mathrm{F}_{\mathrm{fit}}$). The data and corresponding fits obtained for the three flux values $\Phi_{\mathrm{ext}}/\Phi_{0} = 0, 0.2$ and $0.4$ at low drive powers corresponding to the single photon level are plotted in Fig.\,\ref{Fig:Fig4}(e)-(g) in the complex plane. In general, we obtain good agreement between the $S_{21}^\mathrm{MC}$ data and the least-square fits by Eq.~(\ref{ioref}) for the entire flux range $-0.5 \le \Phi_\text{ext}/\Phi_0 \le +0.5$ and a wide range of probe powers corresponding to an average photon number $1 \le \langle n\rangle \le 10^4$ in the resonator. For all flux values, the JPC is in the slightly overcoupled regime, $Q_{\mathrm{ext}} < Q_{\mathrm{int}}$. The dependence of $Q_{\mathrm{int}}$ on the applied flux and probe power is shown in Fig.\,\ref{Fig:Fig5}(a). From our measurements on the JPCs, we extract $Q_{\mathrm{ext}}$ values in the range from $3.9 \times 10^4$ to $4.1 \times 10^4$ and $Q_{\mathrm{int}}$ values in the range from  $1.1 \times 10^5$ to $1.34 \times 10^5$. 

For the JPA devices, the scattering coefficient $S_{21}^\mathrm{M}$ is also measured around the resonant frequency $\omega_{\mathrm{JPA}}$ with unequal frequency spacing [see Fig.\,\ref{Fig:Fig3}(b)]. In this case, we exploit its wide tunability by external flux to measure the background contribution, $S_{21}^\mathrm{B}$. As can be seen in Fig.\,\ref{Fig:Fig3}(b), for an applied flux of $\Phi_{\mathrm{ext}}/\Phi_0 = 0.5$, the resonance frequency drops well below the frequency range of interest. Utilizing this flux point, we first acquire the background contribution $S_{21}^\mathrm{B}$ in the vicinity of the $\omega_{\mathrm{JPA}}(0.5)$. Then, the reflection coefficient, $S_{21}^\mathrm{M}$, is acquired for the actual working point within the frequency range of interest. Normalizing $S_{21}^\mathrm{M}$ by complex division with the background $S_{21}^\mathrm{B}$, we obtain $S_{21}^\mathrm{MC} = S_{21}^\mathrm{M} / S_{21}^\mathrm{B}$. Although this is a straightforward method to correct for the  background contribution of the microwave setup, it does not address all experimental distortions. In particular, a potential displacement of the ideal resonance circle due to a Fano interference may persist \cite{hornibrookSuperconductingResonatorsParasitic2012a,khalilAnalysisMethodAsymmetric2012b}. 

With the near-ideal circle obtained with the background-corrected $S_{21}^\mathrm{MC}$ data, we extract the total loss rate $\kappa$ by shifting the $S_{21}^\mathrm{MC}$ data to the origin of the complex plane to obtain $S_{21}^\mathrm{C}$. Then, $\kappa$ is obtained from a phase angle $\theta$ vs. detuning $\Delta$ fit to the $\angle S_{21}^\mathrm{C}$ according to Eq.~(\ref{pvsf}) . 
Next, we extract $\kappa_{\mathrm{ext}}$ using a constrained minimization of the real and imaginary parts of $S_{21}^\mathrm{MC}$ with Eq.~(\ref{ioref}) as explained below. Our JPA is deliberately designed with a small external quality factor to achieve a relatively large amplification bandwidth, essential for the experimental applications towards the generation of pure and strongly-squeezed propagating microwave states. However, as our JPA is strongly overcoupled, $Q_{\mathrm{int}} \gg Q_{\mathrm{ext}}$, an accurate estimation of $Q_{\mathrm{int}}$ is difficult as the shape of the resonance curve is dominated by the external coupling. Therefore, a series of resonators with varying external coupling strength must be measured to accurately determine $Q_{\mathrm{int}}$. Another alternative is to estimate the magnitude of $Q_{\mathrm{int}}$\cite{mcraeMaterialsLossMeasurements2020}. As all our JPC and JPA samples are fabricated and measured under identical conditions, we assume that the values obtained for the JPC are good first-order approximations for the internal quality factor of the JPA. By relying on this assumption, we constrain the least square fits of the complex scattering coefficient $S_{21}^\mathrm{MC}$ with Eq.~(\ref{ioref}) using the bound $1\times 10^{-5} < Q_{\mathrm{l}}^{-1} - Q_{\mathrm{ext}}^{-1} < 5 \times 10^{-6}$. This approach for estimating $Q_{\mathrm{int}}$ works well. As can be seen in Fig.\,\ref{Fig:Fig4}(f), the fitting curves are in good agreement with the experimental data. Furthermore, for the data taken at the three selected flux values and signal powers corresponding to $\langle n\rangle \thicksim 1$, the estimated quality factors range between $85$ and $150$ for $Q_{\mathrm{ext}}$ and $1.1\times 10^5 $ and $1.5\times 10^5$ for $Q_{\mathrm{int}}$.

In order to further verify the values derived for the internal loss rates of the fabricated JPAs, we use these devices for the generation of squeezed vacuum states, characterized by the squeezing level $S$ and purity $\mu$. The squeezed states are described using the squeezing operator $\hat{S} = \exp ((\xi^*\hat{a}^2 - \xi(\hat{a}^\dagger)^2)/2)$, with $\xi = r e^{i\phi}$ being the complex squeezing amplitude. Here, $r$ denotes the squeezing factor, $\phi$ the squeezing angle in phase space and $\hat{a} = \hat{q} + i\hat{p}$ $(\hat{a}^\dagger = \hat{q} - i\hat{p})$ the photon annihilation (creation) operator with the electromagnetic field quadrature operators $\hat{q}$ and $\hat{p}$, such that $[\hat{q},i\hat{p}] = 1/2$. The degree of squeezing of the quantum state is quantified in decibels as $S = -10 \log_{10} [ \sigma_{\mathrm{s}}^2/0.25 ]$, where $\sigma_{\mathrm{s}}^2$ is the variance of the squeezed quadrature and the vacuum variance is $0.25$. Similarly, the antisqueezing level is $A = 10\log_{10}(\sigma_{\mathrm{as}}^2/0.25)$, with $\sigma_{\mathrm{as}}^2$ denoting the variance of the antisqueezed quadrature. Furthermore, the purity $\mu$ of the squeezed state is defined as $\mu = 1 / (4 \sqrt{\sigma_{\mathrm{s}}^2\sigma_{\mathrm{as}}^2})$. For a completely mixed state, the purity becomes zero, $\mu \xrightarrow[]{}0$, and for an ideal pure state, it is unity, $\mu \xrightarrow[]{}1$.

Using the quantum input-output formalism, we describe the JPA as a nonlinear resonator coupled to two environmental baths, where $\kappa_\mathrm{ext}$ governs coupling to the external circuit and $\kappa_\mathrm{int}$ accounts for internal losses. A complete derivation of this model is provided in the supplementary material of Ref.~\cite{fedorovFinitetimeQuantumEntanglement2018b}. Based on this framework, we obtain the following expressions for the output quadrature variances,
\begin{equation}
\label{variances}
\begin{aligned}
\langle p_\mathrm{b,out}^2 \rangle &= \frac{ (2\chi - \kappa_\mathrm{ext} + \kappa_\mathrm{int})^2 \langle p^2_\mathrm{b,in} \rangle + 4\kappa_\mathrm{ext}\kappa_\mathrm{int} \langle p^2_\mathrm{c,in} \rangle }{(2\chi + \kappa)^2}, \\  
\langle q_\mathrm{b,out}^2 \rangle &= \frac{ (2\chi + \kappa_\mathrm{ext} - \kappa_\mathrm{int})^2 \langle q^2_\mathrm{b,in} \rangle + 4\kappa_\mathrm{ext}\kappa_\mathrm{int} \langle q^2_\mathrm{c,in} \rangle }{(2\chi - \kappa)^2},  
\end{aligned}
\end{equation}
where $\chi = \epsilon \chi^{(2)}$ depends on the pump amplitude $\epsilon$ and the nonlinear susceptibility term $\chi^{(2)}$ \cite{wallsSqueezedStatesLight1983a}. The expectation values $\langle q^2_{\mathrm{b,in}} \rangle$, $\langle p^2_{\mathrm{b,in}} \rangle$, $\langle q^2_{\mathrm{c,in}} \rangle$, and $\langle p^2_{\mathrm{c,in}} \rangle$ represent the quadrature variances associated with the input fields of the two baths. In our analysis, these baths are modeled as thermal states characterized by temperatures $T_{\mathrm{att}}$ and $T_{\mathrm{mxc}}$, leading to the thermal variances 
$
\langle q^2_{b,\mathrm{in}} \rangle = \langle p^2_{b,\mathrm{in}} \rangle = (1 + 2 n_{\mathrm{att}}) / 4, \quad
\langle q^2_{c,\mathrm{in}} \rangle = \langle p^2_{c,\mathrm{in}} \rangle = (1 + 2 n_{\mathrm{mxc}})/4,
$
where $n_{\mathrm{att}}$ and $n_{\mathrm{mxc}}$ denote the thermal photon numbers at frequency $\omega_{\mathrm{JPA}}$. These thermal photons are given by the Planck law, $n = [\exp( \hbar \omega_{\mathrm{JPA}}/k_\mathrm{B}T) - 1]^{-1} $, where $n \in (n_{\mathrm{att}},n_{\mathrm{mxc}})$, $T \in (T_{\mathrm{att}},T_{\mathrm{msc}})$, with $k_\mathrm{B}$ the Boltzmann constant.
Since the degenerate gain $G = e^{2r}$ (or the squeezing level) depends on the pump power, fluctuations in the pump photon number add additional noise to the 
signal \cite{rengerStandardQuantumLimit2021d, kylemarkGainWavelengthDependence2006}. At high pump powers, these fluctuations reduce both the squeezing level and purity of the output state. This effect can be modeled by an additional noise photon number \cite{rengerStandardQuantumLimit2021d}
\begin{equation}
    n_\mathrm{J} = n_\mathrm{J}' (G - 1) ^ \delta ,
\end{equation}
where $n_\mathrm{J}'$ is a constant prefactor, and $\delta$ depends on JPA parameters. This gain-dependent noise modifies the output quadrature variances, leading to a deterioration of squeezing and purity. Assuming the noise contribution is evenly distributed between the squeezed and anti-squeezed quadratures, we obtain the modified output variances for the squeezed and anti-squeezed quadratures
\begin{equation}
    \begin{aligned}
        \sigma_\mathrm{s}^2 &= \langle p_\mathrm{b,out}^2 \rangle + \frac{n_\mathrm{J}}{2}, \\
        \sigma_\mathrm{as}^2 &= \langle q_\mathrm{b,out}^2 \rangle + \frac{n_\mathrm{J}}{2}.
    \end{aligned}
\end{equation}


One important ingredient to improve the purity of a squeezed state generated by the JPA is to decrease losses (increase $Q_{\mathrm{int}}$). Furthermore, the maximum squeezing level that can be generated by a JPA scales inversely with the internal losses, $S_{\mathrm{max}} \propto 1/\kappa_{\mathrm{int}}$, and the dependence of $\mu$ on $S$ is approximately $\mu \approx 1 - (\kappa_{\mathrm{int}} \cdot S)/ \kappa $ \cite{balybinSinglemodeTwomodeSqueezed2020,fedorovFinitetimeQuantumEntanglement2018b,Walls1994}. Therefore, to simultaneously obtain high $S$ and $\mu$, the intrinsic JPA losses are required to be very low. 

\begin{figure}[tb]
	\begin{center}
		\includegraphics[width=\linewidth,angle=0, clip]{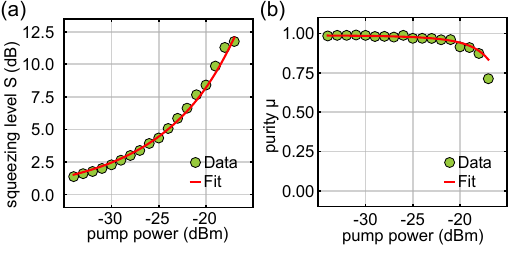}
	\end{center}
	\caption{Squeezing level and purity of squeezed vacuum states generated with the JPA sample. Positive squeezing levels indicate squeezing below vacuum. (a) Experimentally determined squeezing level and (b) purity of squeezed states as a function of the microwave pump power. The value of the pump power is referred to the JPA pump port. }
	\label{Fig:Fig6}
\end{figure}

We generate squeezed vacuum states $\hat{S}(\xi)\ket{0}$ by sending the vacuum state to the JPA operated in the phase-sensitive regime. This is realized by applying a strong coherent microwave pump tone ($\omega_{\mathrm{pump}}/2\pi =2 \omega_{\mathrm{JPA}}/2\pi = \SI{11.08}{\giga\hertz}$) to the JPA pump line. We perform state tomography of propagating squeezed states by using the reference state reconstruction scheme \cite{pogorzalekSecureQuantumRemote2019a,fedorovFinitetimeQuantumEntanglement2018b}. As shown by Fig.\,\ref{Fig:Fig6}, we achieve squeezing levels of up to $S=\SI{11.75}{dB}$ and purities of around $\mu = 98.96\%$.

Using previously determined values of $\kappa_{\mathrm{ext}}$ and experimentally measured temperatures, we simultaneously fit the squeezing and purity data to Eq.~\ref{variances} by minimizing a least-square deviation. The quantities $\kappa_{\mathrm{int}}$, $\chi^{(2)}$, $n_\mathrm{J}'$, and $\delta$ are treated as free fit parameters, while the temperatures $T_{\mathrm{att}}$ and $T_{\mathrm{mxc}}$ are constrained within the experimentally observed range of $\SI{10}{\milli\kelvin}$ to $\SI{50}{\milli\kelvin}$. The resulting fit is shown in Fig.\,\ref{Fig:Fig6} and its parameters yield $T_{\mathrm{att}}=\SI{31}{\milli\kelvin}$ and $T_{\mathrm{mxc}}=\SI{10}{\milli\kelvin}$, along with $\chi^{(2)}= \SI{840}{\mega\hertz}$, $n_\mathrm{J}'=0.0069$, and $\delta=0.047$. Most importantly, we extract the internal quality factor $Q_{\mathrm{int}}=1.26 \times 10^5$, which is in excellent agreement with the estimations derived from the earlier analysis of the scattering data. Although neither $S$ nor $\mu$ have been optimized, these exceptionally high values confirm the low internal loss rates of the realized JPAs.

\medskip

\section{Conclusion}

In this work, we demonstrate the fabrication of high-quality Josephson Parametric Converters (JPCs) and Josephson Parametric Amplifiers (JPAs) with high internal quality factors $Q_{\mathrm{int}}$ in excess of $10^5$ at low signal powers corresponding to the single-photon level. Comparing these $Q_{\mathrm{int}}$ values to those reported in the literature for various values of the coupling efficiency $Q_{\mathrm{ext}}/Q_{\mathrm{l}}$ [cf.~Fig.\,\ref{Fig:Fig5}(b)], we can conclude that, to the best of our knowledge, the values presented in this work are the highest achieved thus far \cite{liFluxCoupledTunable2023a,wangTunableSuperconductingResonators2024,svenssonPeriodtriplingSubharmonicOscillations2017,castellanos-beltranWidelyTunableParametric2007,palacios-laloyTunableResonatorsQuantum2008b,anferovMillimeterWaveFourWaveMixing2020a,yanaiCoherenceNonlinearityMechanical2018c,castellanos-beltranAmplificationSqueezingQuantum2008d,eichlerObservationTwoModeSqueezing2011,krantzInvestigationNonlinearEffects2013b,abdoNondegenerateThreewaveMixing2013,flurinJosephsonMixerSwiss2014,xuMagneticFieldResilientQuantumLimited2023a,simoenCharacterizationMultimodeCoplanar2015,uhlFluxtunableYBa2Cu3O7Quantum2023a,yanaiCoherenceNonlinearityMechanical2018c}. Our low-loss and high-performance devices have been fabricated based on Nb resonators and $\mathrm{Al/AlO_x/Al}$ Josephson junctions galvanically connected with an Al bandaging technique. Our fabrication process combines established surface treatment techniques with optimized argon milling steps. These surface treatment steps improve the quality of the substrate-metal, metal-metal, and metal-air interfaces, thereby minimizing intrinsic loss channels. 

Our results mark a significant step forward in the development of low-loss Josephson parametric devices, which are crucial for quantum-limited amplification, frequency conversion, microwave vacuum squeezing, and entanglement generation, among other tasks. Achieving low losses in related tunable Josephson circuits is essential for advancing quantum technologies, including scalable quantum computing and secure quantum communication, where energy dissipation remains one of the key limitations.

Our results indicate that additional efforts aiming at further improvements in the fabrication routines lead to even lower loss rates and higher $Q_{\mathrm{int}}$ values. In this context, the development of various post-fabrication surface passivation techniques is of particular interest \cite{altoeLocalizationMitigationLoss2022b}. Furthermore, adopting substrate deep trenching methods may allow one to further increase internal quality factors by reducing the lossy Si participation ratio \cite{brunoReducingIntrinsicLoss2015a}. We finally note that expanding these techniques to more complex Josephson parametric devices, such as traveling wave parametric amplifiers (TWPAs), will contribute to the ongoing efforts in building truly quantum-limited broadband parametric amplifiers for scalable quantum computing.

\section{Acknowledgements}

We are grateful to Yuki Nojiri, Florian Fesquet, and Patricia Oehrl for their support and constructive discussions regarding the theoretical aspects of this work. We acknowledge support from the German Research Foundation via Germany’s Excellence Strategy (Grant No. EXC-2111-390814868) and the German Federal Ministry of Education and Research via the QUARATE (Grant No.13N15380) and GeQCoS (Grant No. 13N15680) projects. This research is part of the Munich Quantum Valley, which is supported by the Bavarian state government with funds from the Hightech Agenda Bayern Plus.








\bibliography{Bibliography_JPD}

\clearpage

\end{document}